# A Robust Moment System Based on Absolute Deviations and Quantile Slicing

**Elsayed A. H. Elamir**

Department of Management and Marketing, College of Business Administration, Kingdom of Bahrain

Email: shabib@uob.edu.bh

**Abstract**

This study develops two robust, quantile-sliced moment systems mean and median absolute deviation (MAD and MedAD moments**)** to serve as foundational tools in parametric modelling, statistical inference, describing distributional location, scale, skewness, and tail behaviour in settings where classical moments and L-moments fail. MAD moments use block-wise absolute deviations around the median and exist whenever the mean is finite while MedAD moments replace expectations with medians, ensuring existence for all distributions, including heavy-tailed cases with undefined mean or variance. The systems exhibit strong consistency, slice-based robustness, and bounded influence. The results show that MAD and L-moment ratios are efficient for light to moderate tails, whereas MedAD ratios remain uniquely stable when higher moments do not exist. Application to Cauchy parameter estimation highlight the practical value of MedAD estimators as simple, fully robust alternatives to likelihood-based approach. Together, these systems offer a unified, median-anchored framework for reliable distributional inference under heavy tails and contamination.

# 1 Introduction

Mean absolute deviation (MAD) avoids squaring deviations, which ensures greater stability under heavy-tailed distributions and reduced sensitivity to outliers relative to variance-based measures. Its robustness has led to widespread application across diverse fields; (Gorard, 2005; Pham-Gia and Hung, 2001; Elamir, 2012). In finance, MAD is used in volatility modelling and in mean absolute deviation portfolio optimization, offering risk measures that are less sensitive to extreme returns (Konno et al., 1993; Hosseini et al. 2023). In machine learning, the analogue of MAD is widely used as a loss function and in clustering methods (Shafizadeh et al., 2022; Holste et al. 2025). Across these and other disciplines, the MAD combination of robustness, interpretability, and broad applicability has made it a foundational tool for characterizing dispersion where classical variance-based methods perform poorly, particularly in heavy tailed or contaminated data environments.

Moment-based approach is one of important approaches that estimate distributional characteristics. The oldest and most prominently, the method of moments (MoM, Pearson, 1936) and the more recently developed L-moments (Hosking, 1990). Both frameworks aim to summarize distributional shape, location, scale, skewness, and kurtosis, through functionals computed from data, and both serve as foundational tools in parametric modelling and statistical inference (Hazelton, 2025; Breitung et al., 2022; Shin et al., 2025). The classical method of moments is based on equating population moments

$$\mu_r = \mathrm{E}[(X-\mu)^r], r = 1,2,\ldots$$

with their empirical counterparts and solving the resulting system for the model parameters. Because these moments involve powers of deviations from the mean, MoM is inherently tied to the existence of finite central moments and highly affected by outliers. Even when moments exist, the higher-order terms $(X-\mu)^3$ and $(X-\mu)^4$ amplify the influence of extreme observations. To overcome the instability of central moments, L-moments were introduced as linear combinations of order statistics. The $r$-th population L-moment is given by

$$\lambda_r = \frac{1}{r}\sum_{k=0}^{r-1}(-1)^k \binom{r-1}{k} \mathrm{E}[X_{r-k:r}],$$

where $X_{j:r}$ is the j-th order statistic of a sample of size $r$. The first L-moment is $\lambda_1 = \mathrm{E}[X_{1:1}] = \mu$, This is the population mean from conceptual size 1. The second L-moment is $\lambda_2 = \frac{1}{2}(\mathrm{E}[X_{2:2}] - \mathrm{E}[X_{1:2}])$, this is a robust scale measure and reflects the distance from conceptual size 2. The third L-moments is $\lambda_3 = \frac{1}{3}(\mathrm{E}[X_{3:3}] - 2\mathrm{E}[X_{2:3}] + \mathrm{E}[X_{1:3}])$, these compares left and right outers with median from conceptual size 3. The fourth L-moments is $\lambda_4 = \frac{1}{4}(E[X_{4:4}] - 3E[X_{3:4}] + 3E[X_{2:4}] - E[X_{1:4}])$, this compares two middle parts with two outer parts from conceptual size 4 (Elamir and Seheult, 2003). L-moment estimators exhibit smaller sampling variability and often smaller bias than classical moment estimators, making them effective even when sample sizes are limited. Despite these big improvements, L-moments retain a fatal dependence on the population mean and tail-sensitive (Sankarasubramanian and Srinivasan, 1999; Ulrych et al., 2000; Rychlik and Szymkowiak, 2025).

These limitations leave a methodological gap for characterizing distributional shapes when classical moments and L-moments either do not exist or behave unreliably. This study addresses that gap by developing MAD and MedAD moments, two families of quantile localized absolute deviation-based shape descriptors. Built around the median rather than the mean, these moments remain stable under heavy tails, resistant to outliers, and applicable in settings where standard approaches fail where we establish a comprehensive theoretical framework for these robust alternatives, including their definitions, properties, sampling behaviour, and standardized forms, thereby providing a unified and fully applicable moment system for all distributions, especially those for which conventional methods break down.

This study is organized as follows. Section 2 introduces the proposed MAD moment system. Section 3 presents estimation of MAD moment and its sampling distribution. Section 4 develops the corresponding median absolute deviation moment system. Section 5 presents estimation of median absolute moments; breakdown point and influence curve. Section 6 demonstrates the practical utility of the proposed methods through Cauchy parameter estimation. The conclusion is given in Section 7.

## 2  The proposed MAD Moments

Let $X_1, \ldots, X_n$ be an independent and identically distributed random sample drawn from a continuous population with distribution function $F_X(\cdot)$, where $0 < F_X(x) < 1$, density $f_X(\cdot)$ with $f_X(x) \geq 0$, and quantile function $Q(F)$. Let the population mean be $\mu = E(X)$, the population median be $M = \text{Med}(X)$, and the standard deviation be $\sigma = \sqrt{E[(X-\mu)^2]}$. Denote by $\mathbf{I}_{(i \leq k)}$ the indicator function, which equals 1 if $i \leq k$ and 0 otherwise. The corresponding order statistics of the sample are written as $X_{(1)}, \ldots, X_{(n)}$.

Munoz-Perez and Sanchez-Gomez (1990) provide a key theoretical foundation through the dispersion function $D_X(v) = E|X - v|, v \in R$. Building on this, we introduce a structured, quantile-localized decomposition of the mean absolute deviation $E|X - M|$, evaluated at the

median M. This representation treats the MAD as a special case of the dispersion function and incorporates alternating signs across quantile slices. The $(b+1)^{th}$ MAD moment is defined as

$$\Delta_{b+1} = \begin{cases} M, & b = 0, \\ E|X-M|, & b = 1, \\ \sum_{a=0}^{b-1}(-1)^{a+1}E(|X-M|\ I_{X \in Q(u,v]}), & \text{For } b \geq 2, \end{cases}$$

For continuous distribution

$$\Delta_{b+1} = \begin{cases} F(M) = 0.5, & b = 0, \\ \int_{-\infty}^{\infty}|x-M|f(x)dx, & b = 1, \\ \sum_{a=0}^{b-1}(-1)^{a+1}\int_{Q_u}^{Q_v}|x-M|\ I_{X \in Q(u,v]}f(x)dx, & \text{For } b \geq 2, \end{cases}$$

where M is population median and $I\{\cdot\}$ denotes the indicator function and the term $X \in Q(u,v]$, $u = a/b, v = (a+1)/b$ restricts contributions to observations whose fixed quantiles fall within the corresponding slice of the distribution. This construction parallels to L-moments but replaces linear combinations of order statistics with localized absolute deviations, yielding a family of robust, quantile-anchored measures of location, scale, and higher-order shape. By alternating signs and aggregating across quantile blocks, MAD moments capture subtle distributional asymmetries and peripheral behaviours. When $b \geq 2$ the general form is,

$$\Delta_{b+1} = \sum_{a=0}^{b-1}(-1)^{a+1}E(|X-M|\ I_{X \in Q(u,v]}),$$

This formulation partitions the distribution into $b$ adjacent quantile intervals of equal probability mass, specifically, the intervals $Q(u = a/b, v = (a+1)/b]$ for $a = 0, \ldots, b-1$. Within each slice, the expectation of the absolute deviation $|X-M|$ is computed. By working with absolute deviations around the median, the MAD-moments remain naturally centred on a robust location measure median. Thus, using $|X-M|$ gives MAD moments a clear meaningful interpretation ("average distance from the middle") and ensures that the resulting shape descriptors reflect the dominant structure of the distribution. The first four MAD moments can be written as

$$\Delta_1 = M,$$
$$\Delta_2 = E|X-M|,$$
$$\Delta_3 = -E\left(|X-M|\ I_{X \in Q\left(0,\frac{1}{2}\right]}\right) + E\left(|X-M|\ I_{X \in Q\left(\frac{1}{2},1\right]}\right),$$
$$\Delta_4 = -E\left(|X-M|\ I_{X \in Q\left(0,\frac{1}{3}\right]}\right) + E\left(|X-M|\ I_{X \in Q\left(\frac{1}{3},\frac{2}{3}\right]}\right) - E\left(|X-M|\ I_{X \in Q\left(\frac{2}{3},1\right]}\right),$$

The first MAD moment is median ($\Delta_1 = M$), the well-known a robust measure of central location. The second moment is $\Delta_2 = E|X - M|$, the mean absolute deviation about the median. Unlike the standard deviation $\sigma$, $\Delta_2$ is not inflated by extreme values and is particularly informative for heavy-tailed distributions. The third moment uses two quantile halves

$$\Delta_3 = E\left(|X - M| I_{X \in Q\left(\frac{1}{2}, 1\right]}\right) - E\left(|X - M| I_{X \in Q\left(0, \frac{1}{2}\right]}\right).$$

It compares average absolute deviations below the median to those above it, a distribution with larger deviations on the right side produces $\Delta_3$ (right-skew). A distribution with heavier left-side deviations produces $\Delta_3$ (left-skew). Because the measure relies on absolute deviations rather than cubic terms, $\Delta_3$ provides an interpretable sense of "which side spreads further". The fourth moment divides the distribution into three equal quantile slices

$$\Delta_4 = E\left(|X - M| I_{X \in Q\left(\frac{1}{3}, \frac{2}{3}\right]}\right) - \left(E\left(|X - M| I_{X \in Q\left(0, \frac{1}{3}\right]}\right) + E\left(|X - M| I_{X \in Q\left(\frac{2}{3}, 1\right\}}\right)\right),$$

capturing how strongly the centre slice deviates relative to the peripheral. If $\Delta_4 > 0$, middle deviations dominate (flatter centre / lighter peripheral). If $\Delta_4 < 0$, outer slices dominate. Unlike classical kurtosis, $\Delta_4$ achieves this without cubing, making it robust and more interpretable. It is interesting to note that because the median is the centre of these moments the middle term can be split as

$$E\left(|X - M| I_{X \in Q\left(\frac{1}{3}, \frac{2}{3}\right]}\right) = E\left(|X - M| I_{X \in Q\left(\frac{1}{3}, \frac{1}{2}\right]}\right) + E\left(|X - M| I_{X \in Q\left(\frac{1}{2}, \frac{2}{3}\right]}\right)$$

**Theorem 1.** The MAD moments $\Delta_j$, $j = 2, ...$, of a real valued random variable $X$ exist if and only if $X$ has a finite mean.

*Proof.* From Muñoz-Perez and Sanchez-Gomez (1990) $E|X - v| = v[2F_X(v) - 1] + \mu - 2\int XI_{X \leq v} dF$. When $v = M$, $E|X - M| = \mu - 2\int XI_{X \leq M} dF$ that depend on mean.

A natural extension of the MAD moment framework is obtained by standardizing each higher-order moment $\Delta_{b+1}$ by the robust scale measure $\Delta_2$. The resulting ratio

$$\Gamma_{b+1} = \frac{\Delta_{b+1}}{\Delta_2}, b \geq 2,$$

produces a dimensionless index of distributional shape that is directly comparable across datasets and measurement units. Because $\Delta_2 = E|X - M|$ represents the mean absolute deviation around the median dividing by $\Delta_2$ ensures that $\Gamma_{b+1}$ reflects pure shape information rather than differences in dispersion. For $b = 2$, the standardized skewness measure

$$\Gamma_3 = \frac{\Delta_3}{\Delta_2}$$

compares the magnitude of deviations in the upper half of the distribution to those in the lower half. A positive value indicates greater spread above the median (right-skewness), while a negative value reflects heavier dispersion below the median (left-skewness). For $b = 3$, the standardized measure

$$\Gamma_4 = \frac{\Delta_4}{\Delta_2}$$

assesses the balance of deviations between the centre and peripheral. It can be called peripheral-central measure. If $\Gamma_4 > 0$, deviations in the central slice dominate (flatter centre / lighter periphery), $\Gamma_4 < 0$ deviations in the peripheral slices dominate (more peaked/ heavier-periphery shape), $\Gamma_4 = 0$ peripheral and central deviations are balanced (shape symmetry in dispersion).

**Theorem 2.** Let $X$ be nondegenerate random variable with finite mean. The MAD moments ratio satisfy $|\Gamma_{b+1}| \leq 1$, $b \geq 2$.

*Proof.* Since $\Delta_{b+1}$ is partitions of $\Delta_2 = \sum_{a=0}^{b-1} E\left(|X - M| \; I_{X \in Q\left(\frac{a}{b}, \frac{a+1}{b}\right]}\right)$, with alternating sign, $\Gamma_r$ is bounded.

**Theorem 3.** MAD moments are

(a) location invariance $\Delta_1(X + c) = \Delta_1(X) + c$, $\Delta_b(X + c) = \Delta_b(X)$, $b \geq 2$,

(b) scale equivariance $\Delta_b(aX) = a \Delta_b(X)$, $b \geq 1$.

*Proof.* regarding location invariance, the $\Delta_1(X + c) = \Delta_1(X) + c$, $\Delta_b(X + c) = \Delta_b(X)$, $b \geq 2$, if $Y = X + c$, then $E(Y) = \mu + c$, $|Y - (\mu + c)| = |X - \mu|$. Quantile ranks are unchanged by translation, so slice membership does not change, and thus all $\Delta_b$ for $b \geq 2$. Regarding scale equivariance, for $Y = aX$ with $a > 0$, $E(Y) = a\mu$, $|Y - a\mu| = a|Y - \mu|$. Since mean scale linearly, each term summand in $\Delta_{b+1}$ is times by $a$, yielding $\Delta_b(aX) = a\Delta_b(X)$.

Table 1. MAD moments and standardized shape measures for several distributions.

| Distribution | MAD moments |
|---|---|
| For uniform distribution with parameters $a$ and $b$ $f(x) = 1/(b-a), a < x < b$ | $\Delta_1 = \frac{a+b}{2}, \Delta_2 = \frac{(b-a)}{4},$ $\Gamma_3 = 0, \Gamma_4 = -0.778$ |
| For normal with mean $\mu$ and standard deviation $\sigma$ $f(x) = (1/\sigma\sqrt{2\pi})\exp(-1(x-\mu)^2/2\sigma^2)$ | $\Delta_1 = \mu, \Delta_2 = \sigma\sqrt{2/\pi},$ $\Gamma_3 = 0, \Gamma_4 = -0.823$ |
| For logistic distribution with location $\mu$ and scale $s > 0$, $f(x) = e^{-(x-\mu)/s}/s\left(1+e^{-(x-\mu)/s}\right)^2$ | $\Delta_1 = \mu, \Delta_2 = \sigma\sqrt{2/\pi},$ $\Gamma_3 = 0, \Gamma_4 = -0.837$ |
| For Laplace distribution with location $\mu$ and scale $b>0$, $f(x) = (1/2b)\exp(\|x-\mu\|/b)$ | $\Delta_1 = \mu, \Delta_2 = b$ $\Gamma_3 = 0, \Gamma_4 = -0.875$ |
| For Cauchy with location $\theta$ and scale $s$ $f(x \mid \theta, \gamma) = \frac{1}{\pi s}\left[1+\left(\frac{x-\theta}{s}\right)^2\right]^{-1}$ | $\Delta_1 = \theta, \Delta_2 = \Delta_3 = \Delta_4 = \infty$ |
| Exponential $(\lambda)$ $f(x) = \lambda e^{-\lambda x}, x \geq 0$ | $\Delta_1 = \frac{\ln 2}{\lambda}, \Delta_2 = \frac{1-0.5\ln 2}{\lambda},$ $\Gamma_3 = 0.442, \Gamma_4 = -0.837$ |
| Pareto $f(x) = \frac{\alpha x_m^\alpha}{x^{\alpha+1}}, x \geq x_m$ | $\Delta_1 = x_m 2^{\frac{1}{\alpha}},$ $\Delta_2 = \frac{\alpha x_m}{\alpha-1} - x_m 2^{1/\alpha}$ |

Table 1 shows MAD moments for some distributions. Values of $\Gamma_3$ and $\Gamma_4$ vary according to how strongly peripheral deviations outweigh central ones. Lighter-periphery distributions like the uniform give the least negative $\Gamma_4$, whereas heavier-periphery cases such as the Laplace produce more negative values. The Cauchy has infinite MAD moments beyond $\Delta_1$ due to its extreme heavy-tailed behaviour.

## 3 Estimation of MAD Moments

### 3.1 Sample MAD Moments

Let $x_1, \ldots, x_n$ be an independent sample drawn from a continuous distribution and let $m = \text{median}(x_1, \ldots, x_n)$ denote the sample median. The sample version of the MAD moments provides a fully empirical analogue of the population definitions based on absolute deviations and quantile-slice partitioning. For integers $b \geq 0$, the $(b+1)$-th sample MAD moment is defined as

$$\delta_{b+1} = \begin{cases} m, & b = 0 \\ \dfrac{1}{n}\sum_{i=1}^{n}|x_i - m|, & b = 1 \\ \sum_{a=0}^{b-1}(-1)^{a+1}\dfrac{1}{n}\sum_{i=1}^{n}\left(|x_i - m|\, \mathbf{I}_{x \in q(u,v]}\right), & b \geq 2, \end{cases}$$

where $q(u = a/b, v = (a+1)/b]$ denotes the sample quantile slice corresponding to the interval $(a/b, (a+1)/b]$, $\mathbf{I}\{\cdot\}$ is the indicator function. Thus, for $b \geq 2$, the sample MAD moments are computed by partitioning the sample into $b$ adjacent quantile blocks of equal probability, evaluating the block-wise mean absolute deviations from the sample median, and aggregating these deviations using the alternating sign scheme. As in the population case, the resulting $\delta_{b+1}$ provides a robust measure of distributional shape, with $\delta_1$ estimating the median, $\delta_2$ giving the empirical mean absolute deviation about the median, and higher-order $\delta_{b+1}$ describing asymmetry and peripheral behaviour in a manner resilient to outliers and heavy-tailed observations.

Analogous to the standardized population MAD moments, each higher-order empirical MAD moment may be normalized by the sample scale measure $\delta_2$. For integers $b \geq 2$, the standardized sample MAD moment is defined as

$$\gamma_{b+1} = \frac{\delta_{b+1}}{\delta_2}.$$

Because $\delta_2$ is a robust estimator of scale, dividing by $\delta_2$ produces dimensionless indices of distributional shape that are directly comparable across samples, measurement units, or distinct datasets. The standardized third sample MAD moment

$$\gamma_3 = \frac{\delta_3}{\delta_2}$$

serves as a robust measure of skewness, comparing deviations above and below the sample median. A positive value indicates greater spread in the upper half of the sample, while a negative value indicates heavier dispersion in the lower half. Similarly,

$$\gamma_4 = \frac{\delta_4}{\delta_2}$$

provides a robust measure, reflecting the balance of deviations between the centre portion of the sample and the peripheral. Positive values indicate relatively broader central deviations (flatter centre), while negative values indicate stronger peripheral. Standardization therefore isolates shape features independently of sample scale, preserving the interpretability and robustness of the underlying MAD-based framework. To define a kurtosis measure based on mean absolute deviation, Pinsky (2024) adopts a different approach that depends on the computation of two specific values and the of the building distributions with these reference values as their averages.

## 3.2 Sampling distribution of sample MAD moments

The asymptotic behaviour of the sample MAD moments follows from their representation as linear combinations of block-wise empirical averages of the truncated deviation functions $h_a(X) = |X - M|I\{X \in Q(u,v)\}$. Because each $\bar{h}_a = n^{-1}\sum_{i=1}^{n} h_a(X_i)$ is the sample mean of an i.i.d. sequence, the classical multivariate Central Limit Theorem ensures joint asymptotic normality of $(\bar{h}_0, \ldots, \bar{h}_{b-1})$ around their population expectations. The $(b+1)$-th sample MAD moment is a fixed alternating-sign linear combination of these slice means.

**Theorem 4.** Let
$$h_a(X) = |X - M|\, I\{X \in Q(u,v)\}, a = 0, \ldots, b-1,$$
where M is the population median and $Q(u,v]$ denotes the quantile slice $(u,v]$. Define the sample MAD moment
$$\delta_{b+1} = \sum_{a=0}^{b-1}(-1)^{a+1}\bar{h}_a, \qquad \bar{h}_a = \frac{1}{n}\sum_{i=1}^{n} h_a(X_i),$$
and the corresponding population moment
$$\Delta_{b+1} = \sum_{a=0}^{b-1}(-1)^{a+1}E[h_a(X)].$$
Then, as $n \to \infty$,
$$\sqrt{n}(\delta_{b+1} - \Delta_{b+1}) \xrightarrow{d} N(0, K),$$
where
$$K = \sum_{e=0}^{b-1}\sum_{f=0}^{b-1}(-1)^{e+f} Cov\left(h_e(X), h_f(X)\right)$$

*Proof.*
For each $a \in \{0, \ldots, b-1\}$, define
$$h_a(X) = |X - M|\, I\{X \in Q(u,v)\}$$
The sample counterpart is $\bar{h}_a = \frac{1}{n}\sum_{i=1}^{n} h_a(X_i)$, which is simply the mean of i.i.d. variables $\{h_a(X_i)\}$. Thus, for each fixed $a$,
$$\sqrt{n}(\bar{h}_a - E[h_a(X)])$$

is asymptotically normal by the classical Central Limit Theorem. Because the vector $(\bar{h}_0, \bar{h}_1, \ldots, \bar{h}_{b-1})$ consists of averages of i.i.d. vectors, the multivariate CLT gives

$$\sqrt{n}\big[(\bar{h}_0, \ldots, \bar{h}_{b-1}) - (Eh_0, \ldots, Eh_{b-1})\big] \xrightarrow{d} N(0, \Sigma),$$

where

$$\Sigma = \text{Cov}\big(h_e(X), h_f(X)\big).$$

The sample MAD moment can be written as

$$\delta_{b+1} = \sum_{a=0}^{b-1} (-1)^{a+1} \bar{h}_a.$$

Similarly,

$$\Delta_{b+1} = \sum_{a=0}^{b-1} (-1)^{a+1} E[h_a(X)].$$

Hence the estimation error is

$$\delta_{b+1} - \Delta_{b+1} = \sum_{a=0}^{b-1} (-1)^{a+1} (\bar{h}_a - E[h_a]).$$

Let

$$w = ((-1)^1, (-1)^2, \ldots, (-1)^b)^\top.$$

Then

$$\delta_{b+1} - \Delta_{b+1} = w^\top [(\bar{h}_0, \ldots, \bar{h}_{b-1})^\top - (Eh_0, \ldots, Eh_{b-1})^\top].$$

Because any fixed linear combination of a multivariate normal limit is itself normally distributed,

$$\sqrt{n}(\delta_{b+1} - \Delta_{b+1}) \xrightarrow{d} N(0, w^\top \Sigma w).$$

Since

$$w^\top \Sigma w = \sum_{e=0}^{b-1} \sum_{f=0}^{b-1} (-1)^{e+f} \text{Cov}\big(h_e(X), h_f(X)\big),$$

See, Wasserman (2004).

**Theorem 5.** Let $X_1, \ldots, X_n$ be i.i.d. from a continuous distribution with population median M. Let $\delta_{b+1}$ be the $(b+1)$-th sample MAD moment, defined using the sample median $m$ and empirical quantile slices. If the population MAD moment $\Delta_{b+1}$ is finite, then

$$\delta_{b+1} \xrightarrow{a.s.} \Delta_{b+1} (n \to \infty).$$

*Proof*

Each slice-based term in $\delta_{b+1}$ is an empirical average of the form $(1/n)\sum_{i=1}^{n}|X_i - m|\mathbf{I}\{X_i \in q(u,v]\}$, and converges almost surely to its population expectation because: (i) the sample median $m \to M$ a.s.; (ii) empirical quantiles converge to population quantiles; and (iii) each truncated deviation is integrable. By the strong law of large numbers, each slice average converges a.s. to its expectation. Since $\delta_{b+1}$ is a finite alternating sign linear combination of these terms, it also converges almost surely to

$$\Delta_{b+1} = \sum_{a=0}^{b-1}(-1)^{a+1}E[|X-M|\,\mathbf{I}\{X \in Q(u,v]\}].$$

Hence $\delta_{b+1}$ is strongly consistent. This satisfies to standardized sample MAD moment because $\delta_{b+1} \to \Delta_{b+1}$ a.s., $\delta_2 \to \Delta_2 > 0$ a.s., and the continuous mapping theorem applies (Casella and Berger, 2024; Wasserman, 2004).

## 4 Median Absolute Deviation Moments (MedAD Moments)

A natural and robust extension of the MAD-moment framework can be obtained by replacing the expected value operator in the original definitions with the median functional. This leads to a new family of distributional descriptors that we refer to as Median Absolute Deviation Moments (MedAD-moments). While the classical MAD-moments require the existence of finite expectations, the median exists for every probability distribution on the real line, making the MedAD moment framework applicable to an even broader class of distributions, including those lacking finite mean or finite first absolute moment such as Cauchy distribution (Rousseeuw, and Croux, 1993; Arachchige and Prendergast, 2026; Falk, 1997; Arachchige et al., 2022). Let $M = \text{Med}(X)$ denote the population median. For integers $b \geq 0$, the $(b+1)$-th MedAD-moment is defined as

$$\Phi_{b+1} = \begin{cases} M, & b = 0 \\ \text{Med}|X - M|, & b = 1 \\ \sum_{a=0}^{b-1}(-1)^{a+1}\text{Med}\big(|X-M|\,\mathbf{I}_{X \in Q(u,v]}\big), & \text{For } b \geq 2, \end{cases}$$

where $Q(u = a/b, v = (a+1)/b]$ denotes the quantile slice corresponding to the interval $(u,v]$ of the cumulative distribution function. Thus, for $b \geq 2$, the MedAD moment is obtained by partitioning the support of the distribution into $b$ equiprobable quantile blocks, computing the median absolute deviation within each block, and aggregating these block-wise medians using an alternating sign scheme that parallels the construction of classical MAD moments and L-moments. The structure of the MedAD moments mirrors closely that of the MAD moments defined earlier, with one crucial distinction: the expectation operator $E(\cdot)$ is replaced everywhere by the median functional $\text{Med}(\cdot)$. This substitution significantly increases robustness. Medians are unaffected by extreme values, unlike expectations, and they are well defined even for distributions without finite moments. Consequently, MedAD-moments inherit the spirit of quantile-based shape characterization while avoiding the moment-existence restrictions inherent in the MAD moment framework. The first four MedAD moments can be written as

$$\Phi_1 = M,$$
$$\Phi_2 = \text{Med}|X - M|,$$
$$\Phi_3 = -\text{Med}\left(|X - M|\, I_{X \in Q\left(0, \frac{1}{2}\right]}\right) + \text{Med}\left(|X - M|\, I_{X \in Q\left(\frac{1}{2}, 1\right]}\right),$$
$$\Phi_4 = -\text{Med}\left(|X - M|\, I_{X \in Q\left(0, \frac{1}{3}\right]}\right) + \text{Med}\left(|X - M|\, I_{X \in Q\left(\frac{1}{3}, \frac{2}{3}\right]}\right) - \text{Med}\left(|X - M|\, I_{X \in Q\left(\frac{2}{3}, 1\right]}\right),$$

both systems begin with the population median $\Phi_1 = \Delta_1 = M$. Thus, the two frameworks share the same robust location measure. Second moment $\Phi_2 = \text{Med}|X - M|$ measures the absolute deviation. Because the median is unaffected by extreme outliers, $\Phi_2$ provides an even more robust scale measure and is always well defined, regardless of tail behaviour and well known in literature by median absolute deviation studied by many authors, such as Rasuw and …. (1993). Third moments $\Phi_3$ reflects median imbalance, making it less sensitive to long heavy tail distributions and applicable even when $\Delta_3$ does not exist. Fourth moments $\Phi_4$ measure peripheral-versus-centre deviation using three equal probability blocks yields a measure that is robust to extreme tail mass and well-defined for all distributions. MedAD-moments retain the interpretability and quantile-based decomposition of the original MAD-moments but achieve greater robustness and universal existence. The standardizing MedAD moments are

$$\Psi_{b+1} = \frac{\Phi_{b+1}}{\Phi_2} = \frac{\sum_{a=0}^{b-1}(-1)^{a+1}\text{Med}\left(|X - M|\, I_{X \in Q(u,v]}\right)}{\text{Med}|X - M|}, b \geq 2$$

produces a dimensionless index of distributional shape that is directly comparable across datasets and measurement units. For example, the skewness measure at $b = 2$

$$\Psi_3 = \frac{\Phi_3}{\Phi_2} = \frac{-\text{Med}\left(|X - M|\, I_{X \in Q\left(0, \frac{1}{2}\right]}\right) + \text{Med}\left(|X - M|\, I_{X \in Q\left(\frac{1}{2}, 1\right]}\right)}{\text{Med}|X - M|}$$

and a measure of peripheral centre at $b = 3$

$$\Psi_4 = \frac{\Phi_4}{\Phi_2} = \frac{-\text{Med}\left(|X - M| I_{X \in Q\left(0, \frac{1}{3}\right]}\right) + \text{Med}\left(|X - M| I_{X \in Q\left(\frac{1}{3}, \frac{2}{3}\right]}\right) - \text{Med}\left(|X - M| I_{X \in Q\left(\frac{2}{3}, 1\right]}\right)}{\text{Med}|X - M|}$$

$\Psi_4$ serve as peripheral–central measure, quantifying how absolute deviations in the outer quantile slices compare with those in the central slice. If $\Psi_4 > 0$, deviations in the central slice dominate (lighter periphery), $\Psi_4 < 0$ deviations in the peripheral slices dominate (heavier-periphery shape), $\Psi_4 = 0$ peripheral and central deviations are balanc. Note that the standardizing MedAD moments is unbounded.

Therefore, the MedAD moment system provides a fully robust, quantile-localized hierarchy of location, scale, and higher-order distributional shape descriptors. Because medians are always defined, the MedAD-moment approach extends applicability of absolute deviation-based moment methods to all probability distributions.

**Theorem 6**. Every MedAD-moment $\Phi_{b+1}$ exists and finite for all $b \geq 0$ and for every probability distribution on $\mathbb{R}$.

*Proof.* each MedAD moment $\Phi_{b+1}$ is defined as either the median of $X$, the median of $|X - M|$, or a finite sum of medians of the truncated variables $|X - M| I\{X \in Q(u, v]\}$. Since the median exists for every real-valued distribution, $\Phi_1$ and $\Phi_2$ always exist. For $b \geq 2$, each quantile slice $Q(u, v]$ has positive probability mass $1/b$, so the truncated variable $|X - M|I\{X \in Q(u, v]\}$ is well-defined on a finite interval. Within such a slice, $|X - M|$ is bounded above by the distance between the slice endpoints and the median and therefore has a finite median. Because a finite sum of finite medians is also finite, all MedAD moments $\Phi_{b+1}$ exist and are finite for every distribution.

**Theorem 7.** MedAD moments are

   (c) location invariance $\Phi_1(X + c) = \Phi_1(X) + c, \ \Phi_b(X + c) = \Phi_b(X), b \geq 2$,

   (d) scale equivariance $\Phi_b(aX) = a\Phi_b(X), b \geq 1$.

*Proof.* With respect to location invariance, $\Phi_1(X + c) = \Phi_1(X) + c, \ \Phi_b(X + c) = \Phi_b(X), b \geq 2$, if $Y = X + c$, then $\text{Med}(Y) = M + c, |Y - (M + c)| = |X - M|$. Quantile ranks are unchanged by translation, so slice membership does not change, and thus all $\Phi_b$ for $b \geq 2$. Regarding scale equivariance. For $Y = aX$ with $a > 0$, $\text{Med}(Y) = aM, |Y - aM| = a|Y - M|$. Since medians scale linearly, each summand in $\Phi_{b+1}$ is multiplied by $a$, yielding $\Phi_b(aX) = a\Phi_b(X)$.

Obtaining MedAD moments depends on the distribution of $|X - M|$. Let the random variable

$$Y = |X - M|$$

represents the absolute deviation of $X$ from its median. The distribution of $Y$ plays a central role in MedAD moments, since all higher order deviations are computed through block wise statistics of $Y$ restricted to quantile slices and since $y > 0$, the CDF of $Y$ can be obtained as

$$\Pr(|X - M| \leq y) = \Pr(M - y \leq X \leq M + y).$$

Thus, the distribution function of $Y = |X - M|$ is

$$F_Y(y) = F_X(M + y) - F_X(M - y)$$

This representation is valid for all distributions with a well-defined median. If $X$ admits a density $f$, then $Y$ also has a density given by

$$f_Y(y) = f_X(M + y) + f_X(M - y),$$

If $X$ is symmetric around M, then

$$F(M + y) - F(M - y) = F(M + y) - [1 - F(M + y)] = 2F(M + y) - 1,$$

Hence,

$$F_Y(y) = 2F(M + y) - 1.$$

The density reduces to

$$f_Y(y) = 2f(M + y), \quad y > 0,$$

showing that the distribution of absolute deviations is simply a right-half folding of the parent density around the median.

Therefore, the MedAD moments can be computed as

1. first moment

$$F_X(x_{med}) = 0.5 \text{ or } \Phi_1 = Q_X(0.5).$$

2. Second moment

$$\Phi_2 = Q_Y(0.5) \text{ or } F_Y(y_{med}) = 0.5.$$

3. Higher moments ($b \geq 2$):

    Partition the distribution into $b$ quantile slices

    $$S_a = Q\left(\frac{a}{b}, \frac{a+1}{b}\right), a = 0, \ldots, b - 1.$$

    For each slice, the truncated deviation variable is $Y_a = |X - \Phi_1| \; \mathbb{I}\{X \in S_a\}$.

    Its distribution inside the slice is

    $$F_{Y,a}(y) = F\left(\min\left(\Phi_1 + y, F^{-1}\left(\frac{a+1}{b}\right)\right)\right) - F\left(\max\left(\Phi_1 - y, F^{-1}\left(\frac{a}{b}\right)\right)\right).$$

    The conditional CDF is

    $$\tilde{F}_{Y,a}(y) = b \, F_{Y,a}(y).$$

    The slice median $y_a$ satisfies

$$\tilde{F}_{Y,a}(y_a) = 0.5 \text{ or } y_\alpha = \tilde{Q}_{Y,a}(0.5)$$

The $(b+1)$-th MedAD moment is

$$\Phi_{b+1} = \sum_{a=0}^{b-1}(-1)^{a+1}y_a.$$

**Example**

Uniform Distribution $X \sim U(a,b)$ has $F(x) = \frac{x-a}{b-a}$ and $Q(F) = a + (b-a)F$. Therefore,

1. $\Phi_1 = Q_X(0.5) = \frac{a+b}{2}$,

2. For a uniform distribution, the folded distribution of $|X - M|$ is linear on $\left[0, \frac{b-a}{2}\right]$.
   Solving $F_Y(y) = 0.5$ gives $\Phi_2 = \frac{b-a}{4}$,

3. $\Phi_3$: Partition into $b = 2$ slices $S_0 = Q[a, \Phi_1], S_1 = Q\left(\frac{1}{2}, 1\right] = [\Phi_1, b]$,

   - Truncate Truncated deviations $Y_a = |X - \Phi_1| \cdot 1\{X \in S_a\}$.
   - Slice distributions Both slices are symmetric around $\Phi_1$, so the deviation distribution in each slice is identical. Thus, each slice median is $y_0 = y_1 = \frac{b-a}{4}$.
   - Combine with alternating signs $\Phi_3 = -y_0 + y_1 = -\frac{b-a}{4} + \frac{b-a}{4} = 0$.

4. $\Phi_4$: Partition into $b = 3$ slices $Q\left(0, \frac{1}{3}\right], Q\left(\frac{1}{3}, \frac{2}{3}\right], Q\left(\frac{2}{3}, \frac{3}{3}\right]$. Each slice has width $(b-a)/3$.

   - Slice medians: because the uniform distribution is linear inside each slice, the absolute deviation inside each slice grows linearly from 0 at the median slice boundary. The slice medians are equal in magnitude but follow the alternating-sign pattern: $y_0 = y_2 = \frac{b-a}{6}, y_1 = \frac{b-a}{6}$.
   - Combine $\Phi_4 = -y_0 + y_1 - y_2 = -\frac{b-a}{6} + \frac{b-a}{6} - \frac{b-a}{6} = -\frac{b-a}{6}$.

These match the expected symmetry properties: zero skewness $\Phi_3 = 0$, negative kurtosis-type value $\Phi_4 < 0$, indicating light peripheral s and a flat-top relative to symmetric heavy-tailed distributions.

## 5 Estimation of MedAD moment

### 5.1 Sample MedAD moments

For an independent sample $X_1, \dots, X_n$ drawn from a continuous distribution, the sample MedAD moments provide empirical counterparts to the population quantities defined through quantile-localized medians. Let $m = \text{med}(X_1, \dots, X_n)$ denote the sample median and let the empirical quantiles determine the slice boundaries $q(a/b)$ and $q((a+1)/b)$ for $a = 0, \ldots, b-1$. The $(b+1)$-th sample MedAD moment is then

$$\phi_{b+1} = \begin{cases} m, & b = 0 \\ med|x_i - m|, & b = 1 \\ \sum_{a=0}^{b-1} (-1)^{a+1} med\left(|x_i - m| \, \mathbf{I}_{x_i \in q(u,v]}\right), & \text{For } b \geq 2, \end{cases}$$

Thus, each component is computed as a slice-specific sample median of absolute deviations, and the final estimator is obtained by aggregating these components using the same alternating-sign scheme as in the population definition. Because the sample median and empirical quantiles are strongly consistent, each slice-wise truncated median is a strongly consistent estimator of its population counterpart. Consequently, the sample MedAD moments inherit consistency, and their standardized forms

$$\psi_{b+1} = \frac{\phi_{b+1}}{\phi_2}$$

provide dimensionless, robust estimators of distributional shape. These estimators require no existence of moments and remain well-defined under heavy-tailed or contaminated distributions, reflecting the fully robust nature of the MedAD framework. The statistic

$$\psi_3 = \frac{\phi_3}{\phi_2}$$

captures median imbalance between the upper and lower halves of the distribution, taking positive values when deviations above the median dominate and negative values when lower side deviations are larger. Similarly,

$$\psi_4 = \frac{\phi_4}{\phi_2}$$

compares the typical deviation within the central slice to that in the two outer slices, yielding a robust indicator of peripheral heaviness versus central concentration. Because both measures are based on medians within quantile slices, they remain stable under contamination and heavy peripheral, offering interpretable shape descriptors even when classical moments do not exist.

**Theorem 8.** Let $\Phi_{b+1}$ denote the $(b+1)$-th population MedAD moment and let $\phi_{b+1}$ be the corresponding sample MedAD moment defined using the sample median $m$ and empirical quantile slices. For every $b \geq 0$,

$$\phi_{b+1} \xrightarrow{a.s.} \Phi_{b+1}.$$

That is, sample MedAD moments are strongly consistent estimators of the population MedAD moments.

*Proof.* the sample median $m$ and empirical quantiles converge almost surely to their population counterparts. The truncated variables $|X - m|I\{X_i \in q(u, v)\}$ converge pointwise to the population slice deviations, and the slice-wise sample medians converge almost surely to the population medians. Because each MedAD moment is a finite alternating sum of these slice medians, the sample moment converges almost surely to the population moment. Hence $\phi_{b+1} \xrightarrow{a.s.} \Phi_{b+1}$ for all $b$.

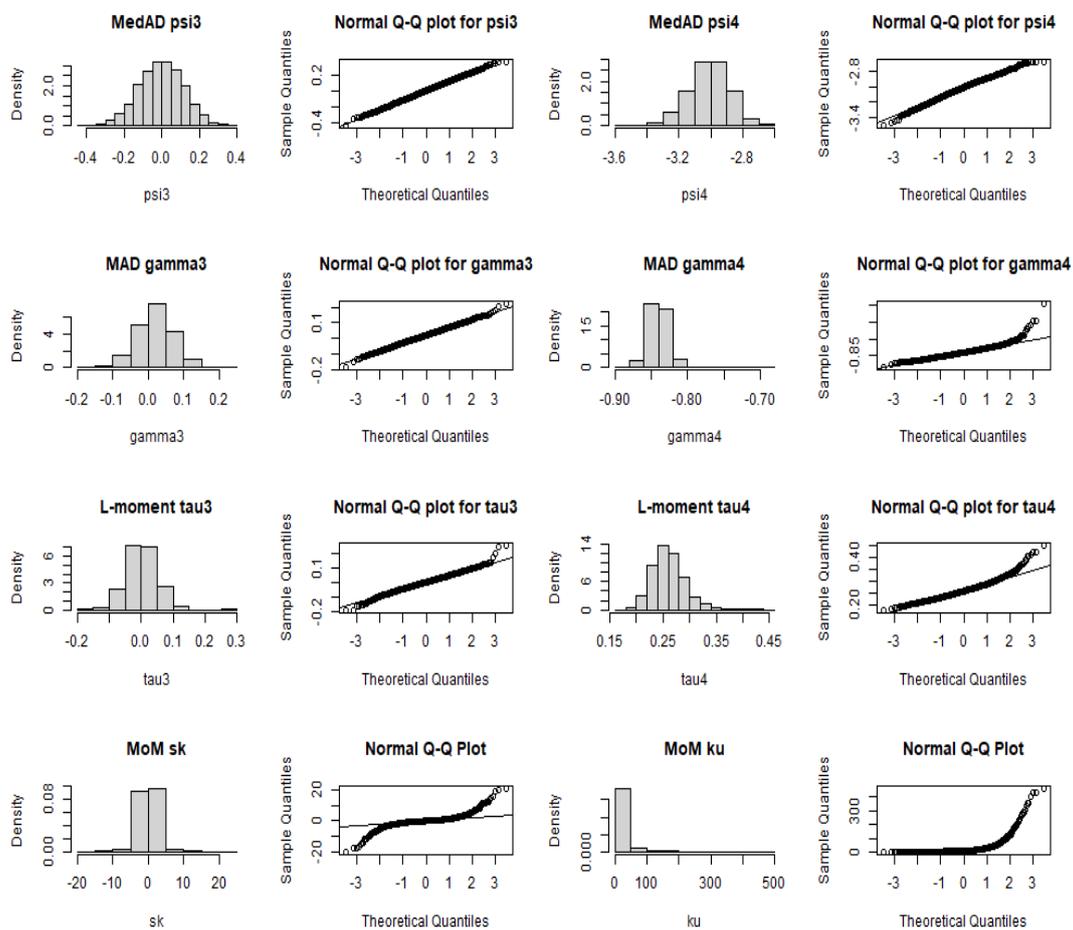

Figure 1. histogram and Q-Q plots for standardized third and fourth moments based on MAD, MedAD, L-moment and MoM moments from $t(df = 3)$ and $n = 500$.

Because the data comes from a t-distribution with $df = 3$, the variance is finite (df > 2) but the distribution still has very heavy tails. With a large sample size ($n = 500$), the stable estimators; particularly the MedAD, MAD, and L-moment ratios; approximate normality for lower-order moments; however, at the fourth order, the estimators $\gamma_4$ and $\tau_4$ become less stable compared with the MedAD-based measure. While L-moment and MAD estimators are tighter than MedAD measures, classical moments suffer from severe distortions, including wide spreads, heavy tails, and curved Q–Q plots.

## 5.2 Breakdown point

For $b \geq 2$, the statistic $\phi_{b+1}$ is a finite alternating sum of $b$ slice-wise medians

$$\phi_{b+1} = \sum_{a=0}^{b-1} (-1)^{a+1} \phi_{b+1,a}, \quad \phi_{b+1,a} = \text{med}(|X_i - m| \; I\{X_i \in q(u,v)\}),$$

The slice contains approximately $n/b$ observations. The median inside this slice breaks when half of the slice values are replaced by arbitrarily large contamination. Thus, the smallest number of contaminated observations needed to make $\phi_{b+1,a}$ diverge is

$$m_a^* = \left\lceil \frac{n_a}{2} \right\rceil = \left\lceil \frac{n}{2b} \right\rceil.$$

Since $\phi_{b+1}$ is an alternating sum of the slice medians, if any slice median $\phi_{b+1,a}$ diverges, then the entire statistic diverges. Thus, the minimal contamination needed to break $\phi_{b+1}$ is exactly the minimal contamination needed to break one slice

$$m^*(\phi_{b+1}) = m_a^* = \left\lceil \frac{n}{2b} \right\rceil.$$

The finite-sample breakdown point (proportion of contaminated data) is

$$\epsilon^*(\phi_{b+1}) = \frac{m^*(\phi_{b+1})}{n} = \frac{1}{2b}.$$

For the first two MedAD moments ($b = 0,1$), which do not use quantile slices, the breakdown point is $\epsilon^*(\phi_1) = \epsilon^*(\phi_2) = \frac{1}{2}$. This result shows the natural trade-off between robustness and resolution. The more slices a MedAD moment uses, the smaller the breakdown point, because each slice median depends on fewer observations (Hampel, 1985; Hekimoglu, 1997).

Figure 2 compares the sampling distributions and normal Q–Q plots of standardized shape estimators under a distribution with undefined variance. Because the variance does not exist, the MAD-, L-moment-, and classical moment–based measures behave poorly and cannot be reliably computed, leaving the MedAD ratios as the only suitable estimators for this setting. This highlights the unique value of MedAD moments; they remain fully defined and applicable even when all other moment-based methods fail.

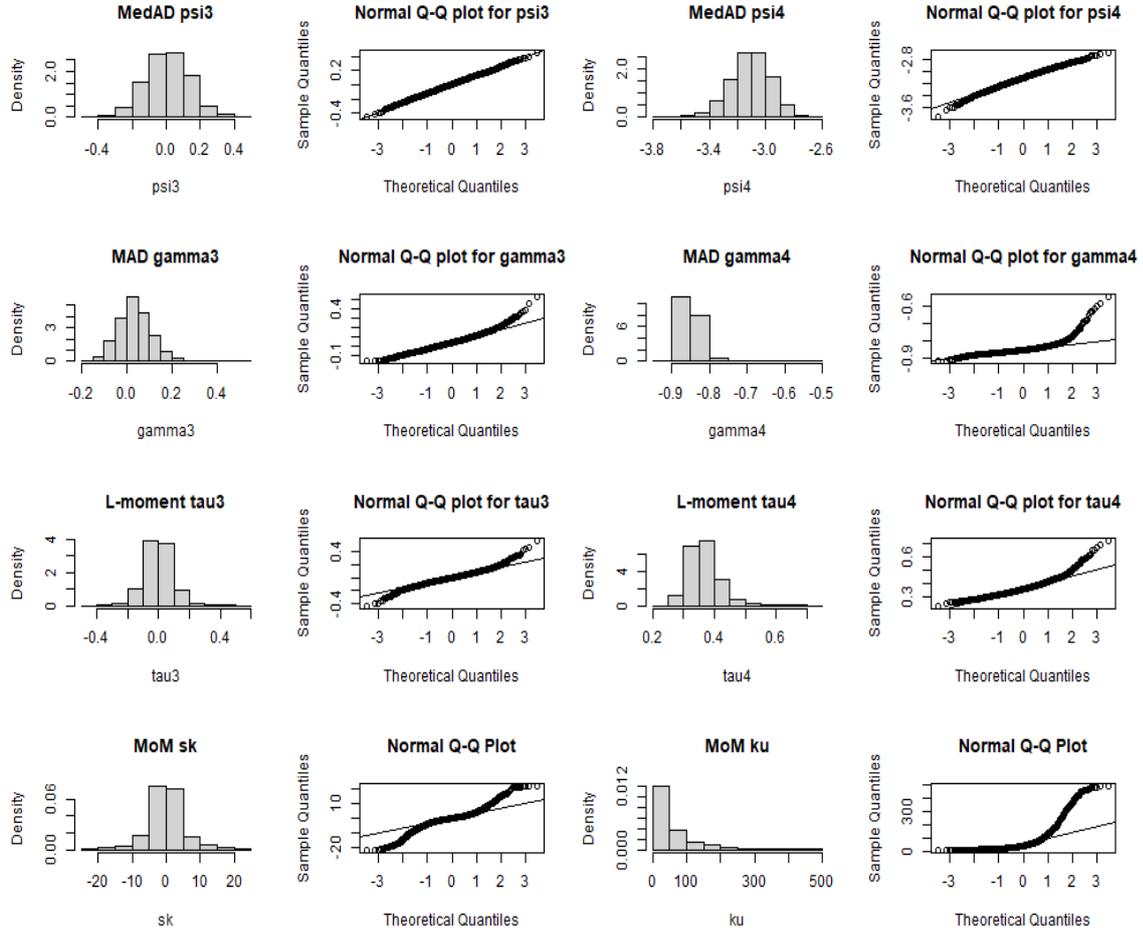

Figure 2. histogram and Q-Q plots for standardized third and fourth moments based on MAD, MedAD, L-moment and MoM moments from $t(df = 2)$ distribution and $n = 500$.

### 5.3 Influence function

Because the influence function is a population concept, we derive the IF for the population MedAD moments $\Phi_{b+1}$, which in turn determines the asymptotic influence of the sample estimator $\phi_{b+1}$. Let $F$ be a distribution with median $M = \text{Med}(F)$. The $(b+1)$-th population MedAD moment is

$$\Phi_{b+1}(F) = \sum_{a=0}^{b-1} (-1)^{a+1} \text{Med}_F(|X - M| \, I\{X \in Q(u, v]\}),$$

where $Q(u, v]$ is a quantile slice of $F$. Let $T(F) = \Phi_{b+1}(F)$. The influence function is

$$\mathrm{IF}(z;T,F) = \lim_{\varepsilon \to 0} \frac{T((1-\varepsilon)F + \varepsilon\Delta_z) - T(F)}{\varepsilon}.$$

(Hampel, 1974; Ruppert, 1987; Hekimoglu, 1997)).

For any scalar functional $m(F) = \mathrm{Med}(F)$, the classical IF is

$$\mathrm{IF}(z;m,F) = \frac{1/2 - I(z \leq M)}{f(M)},$$

where $f(M)$ is the density at the median. For slice $a$, define the slice-restricted deviation variable

$$Y_a = |X - M|\, I\{X \in S_a\},\ S_a = Q(u,v].$$

Let the population slice median be $\theta_a = \mathrm{Med}(Y_a)$. Because $Y_a$ is a univariate variable, its median has influence

$$\mathrm{IF}(z;\theta_a,F) = \frac{1/2 - I(Y_a(z) \leq \theta_a)}{f_{Y_a}(\theta_a)},$$

where

$$Y_a(z) = |X - M|\, I(z \in S_a),$$

and $f_{Y_a}(\cdot)$ is the density of $Y_a$ at $\theta_a$. Since $Y_a$ depends on $M$, the perturbation at point $z$ affects $Y_a$ through $Y_a(X) = |X - M|\, I(X \in S_a)$. Thus the Gateaux derivative adds

$$\frac{\partial Y_a}{\partial M} = -\mathrm{sign}(X - M)\, I(X \in S_a).$$

Therefore, the IF contribution from the shift of the centre $M$ is

$$\mathrm{IF}_M(z;\theta_a,F) = \frac{\partial \theta_a}{\partial M}\bigg|_F \cdot \mathrm{IF}(z;M,F).$$

The influence function for slice $a$ is

$$\mathrm{IF}(z;\theta_a,F) = \frac{1/2 - I(|X - M|\, I(z \in S_a) \leq \theta_a)}{f_{Y_a}(\theta_a)} + \left(\frac{\partial \theta_a}{\partial M}\right) \cdot \frac{1/2 - I(z \leq M)}{f(M)}.$$

The full MedAD moment is a linear combination $\Phi_{b+1}(F) = \sum_{a=0}^{b-1}(-1)^{a+1}\theta_a$. Thus the IF is the same alternating sum of slice IF functions

$$\mathrm{IF}(z;\Phi_{b+1},F) = \sum_{a=0}^{b-1}(-1)^{a+1}\left[\frac{1/2 - I(|z - M|\, I(z \in S_a) \leq \theta_a)}{f_{Y_a}(\theta_a)} + \left(\frac{\partial \theta_a}{\partial M}\right) \cdot \frac{1/2 - I(z \leq M)}{f(M)}\right]$$

Asymptotic Influence of the Sample MedAD Estimator $\phi_{b+1} = \sum_{a=0}^{b-1}(-1)^{a+1}\hat{\theta}_a$, is an empirical plug-in estimator (Law, 1986). By standard M-estimator theory

$$\sqrt{n}(\phi_{b+1} - \Phi_{b+1}) \to_d N(0, \mathrm{E}[\mathrm{IF}(Z;\Phi_{b+1},F)^2]).$$

Then the plug-in estimator of the IF evaluated at a point $z$ is

$$\widehat{\text{IF}}(z; \Phi_{b+1}) = \sum_{a=0}^{b-1}(-1)^{a+1}\left[\frac{1}{2} - \mathbf{I}\left(|z-m|\,\mathbf{I}_{\{z\in S_a\}} \leq \hat{\theta}_a\right)\frac{1}{\hat{f}_{Y_a}(\hat{\theta}_a)} + \frac{\widehat{\partial\theta_a}}{\partial m}\left(\frac{1}{2} - \mathbf{1}(z \leq m)\right)\frac{1}{\hat{f}(m)}\right].$$

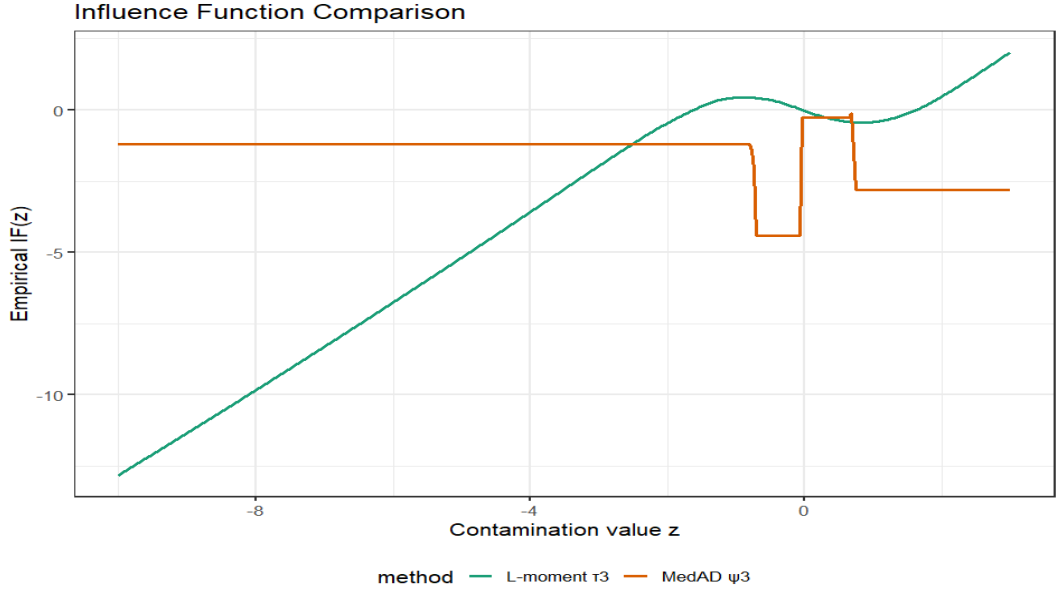

Figure 3. EIF for L-moment $\tau_3$ and $\psi_3$ using data from normal distribution

Figure 3 compares the empirical influence functions of the L-moment skewness estimator $\tau_3$ and the MedAD skewness estimator $\psi_3$ under point contamination. The influence function of $\tau_3$ displays a smooth, continuous, and ultimately unbounded response. As the contaminating value $z$ moves into the extreme tails, the IF grows without limit, indicating that $\tau_3$ remains sensitive to sufficiently large outliers. In contrast, the MedAD estimator $\psi_3$ exhibits a piecewise, step-shaped influence function with clear flat regions and abrupt transitions when $z$ crosses the quantile-slice boundaries used in its construction. Note that R code for computing empirical influence is available from the author upon request.

## 6  Applications

Estimating the parameters of the Cauchy distribution is challenging because its mean and variance do not exist, making classical moment-based methods unusable. As a result, most established approaches rely on likelihood, quantiles, or robust statistics. The MedAD moments for Cauchy distribution is  $\Phi_1 = \theta$, and  $\Phi_2 = s$ Therefore, $\hat{\theta} = \phi_1 = m$, $\hat{s} = \phi_2 = Med|x_i - m|$. To produce the results shown in Table 2, we perform a simulation study comparing four estimators of the Cauchy distribution parameters $\theta$ and $s$: MLE, MedAD, Quantile, and MGOF. The motivation is that the Cauchy distribution lacks a finite mean and variance, making moment-based estimators unusable, while MedAD remains well-defined with $\Psi_1 = \theta$ and $\Psi_2 = s$. For each scenario, samples are drawn from a Cauchy ($\theta$, s) distribution. Three sample sizes are considered; n = 25, 50, 100, as used in Table 2. Estimators Compared are

- MLE: efficient under clean data but sensitive to outliers,
- MedAD: $\hat{\theta}$= sample median, $\hat{s}=\psi_2$, fully robust.
- Quantile estimator: based on sample quantiles,
- MGOF: minimum goodness-of-fit estimator, weaker performance under heavy tail.

For each sample size, generate many (B = 10,000) independent samples from Cauchy ($\theta, s$), All four estimators are computed for each sample. The bias = $E(\hat{\theta}) - \theta$, and $MSE = E(\hat{\theta} - \theta)^2$ and for $s$. Table 2 summarizes the bias and mean squared error (MSE) for estimating the location (θ) and scale ($s$) parameters of the Cauchy distribution using four competing methods (Cane, 1974; McCullagh, 1993; Pekasiewicz, 2014, Delignette-Muller and Dutang, 2015; R Core Team, 2026).

The simulation results clearly demonstrate the classical robustness–efficiency trade-off. MLE remains the most efficient estimator in uncontaminated Cauchy samples, achieving the lowest MSE for both θ and $s$ at all sample sizes. In contrast, MedAD and quantile estimators offer superior robustness at the cost of modest efficiency loss, which is consistent with their median-based construction and bounded influence properties. The MGOF estimator is the least competitive, particularly under small-sample heavy-tailed conditions. These findings reinforce the role of MedAD and similar quantile-based procedures as valuable alternatives when robustness to outliers or contamination is desired, while MLE remains optimal under strict model validity.

Table 2. Bias and MSE for Cauchy parameters estimation based on different methods

|   |          | 25 Bias | 25 MSE | 50 Bias | 50 MSE | 100 Bias | 100 MSE |
|---|----------|---------|--------|---------|--------|----------|---------|
| $\theta$ | MLE      | 0.001   | 0.0895 | -0.003  | 0.0421 | 0.002    | 0.0196  |
|   | MedAD    | 0.004   | 0.1040 | -0.003  | 0.0512 | 0.001    | 0.0245  |
|   | Quantile | 0.004   | 0.1040 | -0.003  | 0.0512 | 0.001    | 0.0245  |
|   | MGOF     | 0.005   | 0.1260 | -0.002  | 0.0589 | 0.002    | 0.0280  |
|   | MedAD    | 0.004   | 0.1040 | -0.003  | 0.0512 | 0.001    | 0.0245  |
| $s$ | MLE      | -0.003  | 0.0926 | 0.000   | 0.0405 | 0.000    | 0.0206  |
|   | MedAD    | 0.026   | 0.1352 | 0.018   | 0.0527 | 0.005    | 0.0262  |
|   | Quantile | 0.026   | 0.1352 | 0.018   | 0.0527 | 0.005    | 0.0262  |
|   | MGOF     | 0.057   | 0.1164 | 0.028   | 0.0471 | 0.014    | 0.0220  |
|   | MedAD    | 0.031   | 0.1212 | 0.017   | 0.0489 | 0.006    | 0.0252  |

# 7 Conclusion

This study develops two quantile-sliced absolute deviation moment systems MAD moments and MedAD moments as robust alternatives to classical moments and L-moments to serve as a tool of parametric modelling and summarizing location, scale, skewness, and tail behaviour. MAD moments retain existence under finite mean and achieve robustness through median and alternating sign block-wise absolute deviation aggregation. Their standardized forms provide dimensionless shape descriptors with strong consistency and asymptotic normal distribution. To address distributions lacking finite first absolute moments, the MedAD system replaces expectations with slice-wise medians, ensuring existence for all distributions, including those with undefined mean or variance. Breakdown-point analysis of MedAD-moments demonstrates a clear robustness resolution trade-off where the first two MedAD moments achieve 50% breakdown point while higher order achieves $1/(2b)$.

The applicability of the MedAD framework is further demonstrated through Cauchy parameter estimation. In this case, the MedAD moments yield simple, fully robust estimators that remain well-defined for all sample sizes and tail configurations. Simulation experiments confirm the robustness–efficiency trade-off where the maximum likelihood estimation achieves lower MSE under uncontaminated data, but MedAD and quantile-based estimators maintain stability and bounded influence under heavy tailed or contaminated samples while minimum goodness-of-fit methods degrade substantially.

Therefore, the MAD and MedAD frameworks form a unified, quantile-structured, median-anchored family of robust moments that remain valid across moment non-existent settings. Future research may extend MAD and MedAD moments to estimation of different distributions such as Pareto and Weibull distributions and compare it with other methods.

**Conflict of interest:** The author does not have any conflict of interest.

**Financial support:** No financial support